\documentclass[journal]{new-aiaa}
\usepackage[utf8]{inputenc}

\usepackage{graphicx}
\usepackage{amsmath}
\usepackage[version=4]{mhchem}
\usepackage{siunitx}
\usepackage{longtable,tabularx}
\usepackage{pstricks}

\hypersetup{bookmarksnumbered, pdfstartview={FitH}, urlcolor=cyan,
 citecolor=blue, menucolor=blue, pagecolor=blue, linkcolor=blue}

\usepackage{amsmath,latexsym,amssymb}
\usepackage{amsfonts,verbatim}
\usepackage{threeparttable,booktabs,dcolumn}
\usepackage{dsfont}

 \newtheorem{thm}{Theorem}[section]

 \newtheorem{problem}{Problem}[section]

 \newtheorem{exam}{Example}[section]

\newtheorem{bcs}{\small}[section]
 \def\pp{{\it Proof.~~}}

 \newcommand{\abs}[1]{\left\vert#1\right\vert}

\usepackage{multirow,booktabs}

\title{Hohmann Transfer via Constrained Optimization}

\author{Li Xie\footnote{Professor,
School of Control and Computer Engineering; the correspondence author, lixie@ncepu.edu.cn
}}
\affil{State Key Laboratory of Alternate Electrical Power System with Renewable Energy Sources\\North China Electric Power University, Beijing 102206, P.R.~China}
\author{Yiqun Zhang\footnote{Senior Research Scientist, yiqunzhang@hotmail.com} and Junyan Xu\footnote{Associate Research Scientist, junyan\_Xu@sina.cn}
}
\affil{Beijing Institute of Electronic Systems Engineering, Beijing 100854, P.R.~China}

\begin{document}

\maketitle

\begin{abstract}
In the first part of this paper, inspired by the geometric method of Jean-Pierre Marec, 
we consider the two-impulse Hohmann transfer problem between two coplanar circular orbits as a constrained nonlinear programming problem. By using the Kuhn-Tucker theorem, we analytically prove the global optimality of the Hohmann transfer. Two sets of feasible solutions are found, one of which corresponding to the Hohmann transfer is the global minimum, and the other is a local minimum. In the second part, we formulate the Hohmann transfer problem as two-point and multi-point boundary-value problems by using the calculus of variations. With the help of the \texttt{Matlab} solver \texttt{bvp4c}, two numerical examples are solved successfully, which verifies that the Hohmann transfer is indeed the solution of these boundary-value problems.  Via static and dynamic constrained optimization,  the solution to  the orbit transfer problem proposed by W. Hohmann ninety-two years ago and its global optimality are re-discovered. \end{abstract}

\section{Introduction}

\lettrine{I}{n} 1925, Dr. Hohmann, a civil engineer published his seminal book \citep{Hohmann} in which he first described the well-known optimal orbit transfer  between two circular coplanar space orbits by numerical examples, and this transfer is now generally called the Hohmann transfer. 
Hohmann claimed that the minimum-fuel impulsive transfer orbit is an elliptic orbit tangent to both  the initial and final circular orbits. 
However a mathematical  proof to its optimality was not addressed until 1960s. The first proof targeted to the global optimality was presented by Barrar  \citep{Barrar} in 1963, where the Whittaker theorem in classical analytical dynamics was introduced and the components of the velocity at any point on an elliptic orbit, perpendicular to its radius vector and the axis of the conic respectively, were used to be coordinates in a plane; see also Prussing and Conway's book \cite[eq.~(3.13)]{Prussing_book} in detail.   
Before that,  Ting in \citep{Ting1960}  obtained the local optimality of the Hohmann transfer. 
 By using a variational method, Lawden investigated the optimal control problem of a spacecraft in an inverse square law field, and invented the prime vector methodology \citep{Lawden_book}.  According to different thrusts,  the trajectory of a spacecraft is divided into the arcs of three types: (1) null thrust arc; (2) maximum thrust arc; (3) intermediate arc. If we approximate the maximum thrust by an impulse thrust, then the orbit transfer problem can be studied by the the prime vector theory and the Hohmann transfer is a special case, which can also be seen in the second part of this paper. Lawden derived all necessary conditions that the primer vector must satisfy. A systemic design instruction can be found in \citep{prussing_2010}. 
 
In 2009, Pontani in \citep{Pontani2009} reviewed the literature concerning the optimality of the Hohmann transfer during 1960s and 1990s, for example \cite{Moyer1965,Battin_book,Marec_book,Hazelrigg1984,Palmore1984,Prussing1992,Yuan1995}. Among of them, Moyer verified the techniques devised respectively  by Breakwell and Constensou via variational methods for general space orbit transfers. The dynamical equations involved were established for orbital elements. 
Battin and Marec used a Lagrange multiplier method and a geometric method in light of a hodograph plane, respectively. 
It is Marec's method that enlightens us to consider the Hohmann transfer in a different way. 
Based on  Green's theorem, Hazelrigg established the global optimal impulsive transfers. Palmore provided an elemental proof with the help of the gradient of the characteristic velocity, and Prussing simplified Palmore's method by utilizing the partial derivatives of the characteristic velocity, and a similar argument also appeared in \cite{Vertregt} which was summarized in the book \cite{Cornelisse_book};  
see more work of Prussing in \cite[Chapter 6]{Prussing_book}. Yuan and Matsushima carefully made use of two lower bounds of velocity changes for orbit transfers and showed that the Hohmann transfer is optimal both in total velocity change and each velocity change.  

Recently, Gurfil and Seidelmann in their new book \citep[Chapter 15]{Gurfil_book} consider the effects of the Earth's oblateness $J_2$ on the Hohmann transfer. Then the velocity of a circular orbit is calculated by  
\[
v=\sqrt{\dfrac{\mu}{r}\left(1+\dfrac{3J_2r_{eq}^2}{2r^2}\right)}
\]
 By using Lagrange multipliers, the optimal impulsive transfer is derived. This transfer is referred as to the extended Hohmann transfer which degenerates  to the standard Hohmann transfer as $J_2=0$. 
Avenda{\~n}o {\em et al.}~in \citep{Avendano}
present a pure algebraic approach to the minimum-cost multi-impulse orbit-transfer problem. 
By using Lagrange multipliers, as a particular example, the optimality of the Hohmann transfer is also provided by this algebraic approach. These authors are all devoted to proving the global optimality of the Hohmann transfer. 

In the first part of this paper, we present a different method to prove the global optimality of the Hohmann transfer. Inspired by the geometric method of Marec in \citep[pp. 21-32]{Marec_book}, we transform the Hohmann transfer problem into a constrained nonlinear programming problem, and then by using the results in nonlinear programming such as the well-known Kuhn-Tucker theorem, we analytically prove the global optimality of the Hohmann transfer. 
Here by the global optimality, we mean that the Hohmann transfer is optimum among all possible two-impulse coplanar transfers. Two sets of feasible solutions are found, one of which corresponding to the Hohmann transfer is the global minimum, and the other is a local minimum.  
In the second part of the paper, 
we consider the Hohmann transfer problem as dynamic optimization problems.  By using variational method, we first present all necessary conditions for two-point and multi-point boundary-value problems related to such dynamic optimization problems, and we then solve two numerical examples with the help of  \texttt{Matlab} solver \texttt{bvp4c}, which verifies that the Hohmann transfer is indeed the solution of these constrained dynamic optimization problems.  By formulating the Hohmann transfer problem as constrained optimization problems,  the solution to the orbit transfer problem proposed by W. Hohmann 92 years ago and its global optimality are re-discovered. 

\section{The optimality of the Hohmann Transfer} \label{sect_2_2}

The Hohmann transfer is a two-impulse orbital transfer from one circular orbit to another; for the background see, e.g., \cite{Curits_book, Longuski_book, Marec_book, Prussing_book}. 
We use the same notation as Marec in \citep[pp. 21-32]{Marec_book}. Let $O_0$ denote the initial circular orbit of a spacecraft, and the radius and the velocity of the initial circular orbit are $r_0$ and $V_0$ respectively. Let $O_f$ denote the final circular orbit, and its radius and velocity are $r_f$ and $V_f$ respectively. For simplicity, let $r_f$ be greater than $r_0$. We use  boldface to denote vectors.
The center of these two coplanar circular orbits is at the origin of the Cartesian inertia reference coordinate system. Suppose all transfer orbits are coplanar to $O_0$. Hence we need only to consider the $x-y$ plane of the reference coordinate system. 
Assume that the spacecraft is initially located at the point $(r_0, 0)$ of the initial circular orbit. 
At the initial time $t_0$, the first velocity impulse vector $\Delta \mathbf V_0$ is applied, and its components in the direction of the radius and the direction perpendicular to the radius are  $\Delta X_0, \Delta Y_0$ respectively. Similarly, at the final time $t_f$, the second velocity impulse  $\Delta \mathbf V_f$ occurs, and its components are $\Delta X_f, \Delta Y_f$.  Hence at the time $t_0^+$ just after $t_0$, the components of the velocity of the spacecraft are $(\Delta X_0, V_0+\Delta Y_0)$. 
Then in order to enter the final circular orbit, the components of the velocity of the spacecraft must be $(-\Delta X_f, V_f-\Delta Y_f)$ at the time $t_f^-$ just before $t_f$.

During the period from $t_0^+$ to $t_f^-$, the spacecraft is on a transfer orbit (conic) and the angular momentum and the energy (per unit mass) are conserved, hence we have 
\begin{align}
h&=r_0(V_0+\Delta Y_0)=r_f(V_f-\Delta Y_f)\notag\\
\mathcal E&=\dfrac{1}{2}\left[(V_0+\Delta Y_0)^2+\Delta X_0^2\right]-\dfrac{\mu}{r_0}=\dfrac{1}{2}\left[(V_f-\Delta Y_f)^2+\Delta X_f^2\right]-\dfrac{\mu}{r_f}\label{eq_Aug4_1}
\end{align}
Notice that  for a circular orbit, its radius and velocity satisfy  $V=\sqrt{\dfrac{\mu}{r}}$, $\mu$ is gravitational constant. It follow from \eqref{eq_Aug4_1} that we can express $\Delta X_0, \Delta Y_0$ in terms of  $\Delta X_f, \Delta Y_f$ and vice versa.
In order to simplify equations, at first, using the initial orbit radius and velocity as reference values, we define new variables:
\begin{align}\label{eq_Nov11_01}
\bar{r}_f=\dfrac{r_f}{r_0},\quad v_f=\dfrac{V_f}{V_0}, \quad y_0=\dfrac{\Delta Y_0}{V_0}, \quad x_0=\dfrac{\Delta X_0}{V_0}, \quad y_f=\dfrac{\Delta Y_f}{V_0}, \quad x_f=\dfrac{\Delta X_f}{V_0}
\end{align}
Substituting \eqref{eq_Nov11_01} into \eqref{eq_Aug4_1} yields the following non-dimensional  angular momentum and energy equalities
\begin{align}
h&=(1+y_0)=\bar{r}_f({v}_f-y_f)\notag\\
\mathcal E&=\dfrac{1}{2}\left[(1+y_0)^2+x_0^2\right]-1=\dfrac{1}{2}\left[(v_f-y_f)^2+x_f^2\right]-\dfrac{1}{\bar r_f}
\end{align}
see \citep[p.22]{Marec_book}. Then we express $x_f, y_f$ 
in terms of $x_0, y_0$
\begin{align}
y_f&=v_f-(1+y_0)\bar{r}_f^{-1}=\bar{r}_f^{-1/2}-(1+y_0)\bar{r}_f^{-1}\label{eq_Aug4_02}\\
x_f^2&=x_0^2+(1+y_0)^2-2 (1-\bar{r}_f^{-1})-(v_f-y_f)^2\notag\\&=x_0^2+(1+y_0)^2(1-\bar r_f^{-2})-2 (1-\bar{r}_f^{-1})\label{eq_Aug4_02a}
\end{align}
from which we have 
\begin{align}
\Delta v_f^2&=\left(\dfrac{V_f}{v_0}\right)^2=x_f^2+y_f^2\notag\\
&=x_0^2+(1+y_0)^2(1-\bar{r}_f^{-2})-2 (1-\bar r_f^{-1})+\left(\bar r_f^{-1/2}-(1+y_0)\bar r_f^{-1}\right)^2\notag\\
&=x_0^2+\left(y_0+1-\bar r_f^{-3/2}\right)^2-(\bar r_f-1)(2 \bar r_f^2-\bar r_f-1)\bar r_f^{-3}\notag\\
&=x_0^2+\left(y_0+1-\bar r_f^{-3/2}\right)^2-(\bar r_f-1)^2(2 \bar r_f+1)\bar r_f^{-3}\notag\\
\Delta v_0^2&=x_0^2+y_0^2 \notag
\end{align}
where we define $\Delta v_0=\sqrt{x_0^2+y_0^2}>0$ and $\Delta v_f=\sqrt{x_f^2+y_f^2}>0$.
Thus the cost functional (i.e., the characteristic  velocity) can be written as 
\begin{align}
\Delta v(x_0, y_0)&= \Delta v_0+\Delta v_f\notag\\&=\sqrt{x_0^2+y_0^2}+\sqrt{x_0^2+\left(y_0+1-\bar r_f^{-3/2}\right)^2-(\bar r_f-1)^2(2\bar r_f+1)\bar r_f^{-3}}\notag
\end{align}
Observing \eqref{eq_Aug4_02a}, it is noted that we must make the following constraint on the first impulse 
\begin{align}\label{eq_Nov9_01}
x_0^2+(1+y_0)^2(1-\bar r_f^{-2})-2(1-\bar{r}_f^{-1})\geq 0
\end{align}
That is, when we use the energy conservation to calculate the non-dimensional component of the second impulse $x_f$,  the inequality \eqref{eq_Nov9_01} must hold such that 
a non-negative number is assigned to $x_f^2$, which actually requires that the transfer orbit intersects the inner circle and the outer circle. 

The above  background can be found 
in \citep[Section 2.2]{Marec_book} which is specified here.
Marec used the independent variables $x_0, y_0$ as coordinates in a hodograph plan and in geometric language, an elegant and simple proof was given to show the optimality of the Hohmann transfer. 
Marec's geometric method enlightens us to consider the Hohmann transfer in a different way. Then the global optimality of the Hohmann transfer analytically appears. 

We now formulate the Hohmann transfer problem as a nonlinear programming problem subject to the inequality constraint \eqref{eq_Nov9_01} in which $x_0$ and $y_0$ are independent variables.
\begin{thm}\label{thm01}
The classical Hohmann transfer is the solution of the following constrained optimization problem
\begin{align}\label{eq_Nov12_02}
\min_{x_0, y_0}&~\Delta v(x_0, y_0)\\
\text{s.t.}  & \quad x_0^2+(1+y_0)^2(1-\bar r_f^{-2})-2(1-\bar{r}_f^{-1})\geq 0\notag
\end{align}
Also this solution is the global minimum. 
\end{thm}
\pp
We say that the Hohmann transfer is the solution to the above optimization problem if the pair $(x_0, y_0)$ corresponding to the Hohmann transfer is feasible to \eqref{eq_Nov12_02}. 
As usual, we use Lagrange multiplier method and define the Lagrangian function
as follows 
\begin{align}
F(x_0, y_0)=\Delta v(x_0, y_0)+\lambda \left(
-x_0^2-(1+y_0)^2(1-\bar r_f^{-2})+2 (1-\bar{r}_f^{-1})\right)\label{eq_Aug10_01}
\end{align}
In view of the Kuhn-Tucker theorem, a local optimum of $\Delta v(x_0,y_0)$ 
must satisfy the following necessary conditions
\begin{align}
&\dfrac{\partial F}{\partial x_0}=\dfrac{x_0}{\Delta v_0}+\dfrac{x_0}{\Delta v_f}-2\lambda x_0=0 \label{eq_Aug4_03}\\
&\dfrac{\partial F}{\partial y_0}=\dfrac{y_0}{\Delta v_0}+\dfrac{y_0+1-\bar r_f^{-3/2}}{\Delta v_f}-2\lambda (1+y_0)(1-\bar r_f^{-2})=0\label{eq_Aug6_01}\\
&\lambda \left(
-x_0^2-(1+y_0)^2(1-\bar r_f^{-2})+2 (1-\bar{r}_f^{-1}) \right)=0\label{eq_Aug7_07}\\
&-x_0^2-(1+y_0)^2(1-\bar r_f^{-2})+2 (1-\bar{r}_f^{-1}) \leq 0 \\
&\lambda \geq 0
\end{align}
Notice that we have assumed that $\Delta v_0, \Delta v_f>0$. 
In order to find a feasible solution,  we divide the proof of the first part of this theorem into two cases.

(i) If $\lambda=0$, then 
due to the equations \eqref{eq_Aug4_03} and \eqref{eq_Aug6_01}, we have $ x_0=0$ and 
 \begin{align}
\dfrac{y_0}{\Delta v_0}+\dfrac{y_0+1-\bar r_f^{-3/2}}{\Delta v_f}=0\label{eq_Aug6_02}
 \end{align}
 respectively. We claim that the equality \eqref{eq_Aug6_02} does not hold because 
the equality $x_0=0$ implies that 
\[
\dfrac{y_0}{\Delta v_0}=\dfrac{y_0}{\abs{y_0}}=\pm 1
\]
which contradicts 
\begin{align}
1<\dfrac{y_0+1-\bar r_f^{-3/2}}{\Delta v_f}=
\dfrac{y_0+1-\bar r_f^{-3/2}}{
\sqrt{\left(y_0+1-\bar r_f^{-3/2}\right)^2-(\bar r_f-1)^2(2 \bar r_f+1)\bar r_f^{-3}}
}, \quad \text{or} \quad
\dfrac{y_0+1-\bar r_f^{-3/2}}{\Delta v_f}
<-1 \notag
\end{align}
Therefore the equation \eqref{eq_Aug6_02} has no solution under the assumption $\lambda=0$.

(ii) We now consider the case $\lambda >0$. 
Then the equation \eqref{eq_Aug7_07} yields 
 \begin{align}
-x_0^2 -(1+y_0)^2(1-\bar r_f^{-2})+2(1-\bar{r}_f^{-1})=0 \label{eq_Aug8_06}
 \end{align}
which together with  \eqref{eq_Aug4_02a} leads to $x_f^2=0$. Thus 
\begin{align}
\Delta v_f=\sqrt{x_f^2+y_f^2}=\sqrt{y_f^2}=\abs{y_f}\label{eq_Nov13_06}
\end{align}
and also
 \begin{align}
x_0^2=-(1+y_0)^2(1-\bar r_f^{-2})+2(1-\bar{r}_f^{-1})\label{eq_Aug8_06a}
 \end{align}
Then we obtain $ \Delta v_0^2$ as follows 
 \begin{align}
 \Delta v_0^2=
x_0^2+y_0^2
&=\bar r_f^{-2}(1+y_0)^2-(1+2y_0)+2(1-\bar{r}_f^{-1}) \notag\\
&=\left(\bar r_f^{-1}(1+y_0)-\bar r_f\right)^2+2(1+y_0)-\bar r_f^2-(1+2y_0)+2(1-\bar{r}_f^{-1}) \notag\\
&=\left(\bar r_f^{-1}(1+y_0)-\bar r_f\right)^2+3-\bar r_f^2-2\bar{r}_f^{-1}\label{eq_Aug6_4}
 \end{align}
Thanks to \eqref{eq_Aug4_03}
\begin{align}
\dfrac{x_0}{\Delta v_0}+\dfrac{x_0}{\Delta v_f}-2\lambda x_0=0\label{eq_Nov12_03}
\end{align}
We now show that $x_0=0$ by contradiction. If $x_0\neq 0$, then the equality \eqref{eq_Nov12_03} implies that 
\begin{align}
2\lambda=\dfrac{1}{\Delta v_0}+\dfrac{1}{\Delta v_f}\label{eq_Nov12_04}
\end{align}
Substituting \eqref{eq_Nov12_04} into  \eqref{eq_Aug6_01} gives
\begin{align}
\dfrac{y_0}{\Delta v_0}+\dfrac{y_0+1-\bar r_f^{-3/2}}{\Delta v_f} =(1+y_0)(1-\bar r_f^{-2})\left(\dfrac{1}{\Delta v_0}+\dfrac{1}{\Delta v_f}\right)\notag
\end{align}
Arranging the above equation and using \eqref{eq_Aug4_02} and \eqref{eq_Nov13_06},  we get
\begin{align}
\dfrac{y_0-(1+y_0)(1-\bar r_f^{-2})}{\Delta v_0}&=
\dfrac{\bar r_f^{-3/2}-(y_0+1)\bar r_f^{-2}}{\Delta v_f}= \dfrac{\bar r_f^{-1/2}-(y_0+1)\bar r_f^{-1}}{\Delta v_f}\bar r_f^{-1}=\dfrac{y_f}{\abs{y_f}}\bar r_f^{-1}\notag
\end{align}
where $y_f\neq 0$ since we have assumed $v_f>0$ and just concluded $x_f=0$.
 By further rearranging the above equation, we have
\begin{align}
\dfrac{\bar r_f^{-1}(1+y_0)-\bar r_f}{\Delta v_0}=\dfrac{y_f}{\abs{y_f}}
=\begin{cases}
1, &\text{if $y_f>0$}\\
-1, & \text{if $y_f<0$}
\end{cases}\label{eq_Aug8_05}
\end{align}
It is noted that \eqref{eq_Aug6_4}  gives  a strict inequality 
\begin{align}
 \Delta v_0^2=\left(\bar r_f^{-1}(1+y_0)-\bar r_f\right)^2+3-\bar r_f^2-2\bar{r}_f^{-1}<\left(\bar r_f^{-1}(1+y_0)-\bar r_f\right)^2 \label{eq_Nov13_08a}
\end{align}
where the function $3-\bar r_f^2-2\bar{r}_f^{-1}<0$ since it is decreasing with respect to $\bar r_f$ and $\bar r_f>1$. 
Hence \eqref{eq_Aug8_05} contradicts to \eqref{eq_Nov13_08a} and does not hold, which implies that the assumption $\lambda >0$ and the equality  $x_0\neq 0$ do not hold simultaneously. Therefore $\lambda >0$ leads to $x_f=0$ and $x_0=0$.

Summarizing the two cases above, we now conclude that a feasible solution to \eqref{eq_Nov12_02} must have  
$\lambda^*>0$, $x_f^*=0$ and $x_0^*=0$. Then 
the corresponding normal component $y_0$ can be solved from \eqref{eq_Aug8_06}
\begin{align}
y_{0}^*=\sqrt{\dfrac{2\bar r_f}{1+\bar r_f}}-1, \quad \hat y_0^*=-\sqrt{\dfrac{2\bar r_f}{1+\bar r_f}}-1
\end{align}
Substituting them into \eqref{eq_Aug4_02} and \eqref{eq_Aug6_01} yields 
\begin{align}
y_f^*&=\bar r_f^{-1/2}\left(1-\sqrt{\dfrac{2}{1+\bar r_f}}\right), \quad 
 \lambda^*= 
\dfrac{1}{2(1+y_0^*)(1-\bar r_f^{-2})}\left(1+\dfrac{y_0^*+1-\bar r_f^{-3/2}}{y_f^*}\right)>0\notag\\ 
\hat y_{f}^*&=\bar r_f^{-1/2}\left(1+\sqrt{\dfrac{2}{1+\bar r_f}}\right), \quad
\hat \lambda^*= 
\dfrac{1}{2(1+\hat y_{0}^*)(1-\bar r_f^{-2})}\left(1+\dfrac{\hat y_{0 }^*+1-\bar r_f^{-3/2}}{\hat y_{f }^*}\right)>0 \label{eq_Nov13_09}
\end{align}
One can see that there exist two sets of feasible solutions, one of which $(x_0^*, y_0^*)$ is corresponding to the Hohmann transfer,  and its the Lagrange multiplier is $\lambda^*$.

We are now in a position to show that the Hohmann transfer is the global minimum. We first show that $(x_0^*, y_0^*)$ is  a strict local minimum. 
Theorem 3.11 in \citep{Avriel} gives a second order sufficient condition 
for a strict local minimum.
To apply it, we need to calculate the Hessian matrix of the Lagrangian function $F$ defined by \eqref{eq_Aug10_01} at $(x_0^*, y_0^*)$ 
\begin{align}
\nabla^2 F(x_0^*,y_0^*,\lambda^*)=\begin{pmatrix}
\dfrac{\partial^2 F}{\partial x_0^2} &
\dfrac{\partial^2 F}{\partial x_0\partial y_0} \vspace*{.2cm}\\
\dfrac{\partial^2 F}{\partial y_0\partial x_0} &
\dfrac{\partial^2 F}{\partial y_0^2} 
\end{pmatrix}_{(x_0^*, y_0^*)}
\end{align}
After a straightforward calculation, by the equations \eqref{eq_Aug4_03} and \eqref{eq_Aug6_01} and using the fact $x_0^*=0$, we have 
\begin{align}
\dfrac{\partial^2 F}{\partial x_0\partial y_0}(x_0^*, y_0^*)&=\dfrac{\partial^2 F}{\partial y_0\partial x_0}(x_0^*, y_0^*)=0 \notag
\end{align}
and also
\begin{align}
\dfrac{\partial^2 F}{\partial x_0^2}(x_0^*, y_0^*)&=
\dfrac{1}{y_0^*}+\dfrac{1}{y_f^*}-2\lambda^* =\dfrac{1+\bar r_f^{-1}(1+y_0^*)}{y_0^*(1+y_0^*)(1+\bar r_f^{-1})}>0 \label{eq_Aug28_01}\\
\dfrac{\partial^2 F}{\partial y_0^2}(x_0^*, y_0^*)&=
\dfrac{-b}{\left ((y_0^*+a)^2-b\right )^{3/2}}-2\lambda^*(1-\bar r_f^{-2}) <0 
\end{align}
where $a=1-\bar r_{f}^{-3/2}, b=(\bar r_f-1)^2(2\bar r_f+1)\bar r_f^{-3}>0$. 
Then one can see that the Hessian matrix is an indefinite matrix. Thus we cannot use the positive definiteness of the Hessian matrix, that is, 
\[
z^\text{T}\nabla^2 F(x_0^*,y_0^*,\lambda^*)z>0, \quad \forall
 z\neq 0
\]
as a sufficient condition to justify the local minimum of $(x_0^*, y_0^*)$ as usual.
Fortunately, Theorem 3.11 in \citep{Avriel} tells us that for the case $\lambda^*>0$, that is, 
the inequality constraint is active, when we use the positive definiteness of the Hessian matrix to justify the local minimum, we need only to consider the positive definiteness of the first block of the Hessian matrix with the non-zero vector $z\neq 0$ defined by  
\begin{align}
z\in Z(x_0^*, y_0^*)=\{z: z^{\text T}\nabla g(x_0^*, y_0^*)=0\}\label{eq_Aug8_01}
\end{align}
where $g(x_0,y_0)$ is the constraint function
\[g(x_0,y_0)=-x_0^2-(1+y_0)^2(1-\bar r_f^{-2})+2 (1-\bar{r}_f^{-1})\]
With this kind of $z$, if $z^\text{T}\nabla^2 F(x_0^*,y_0^*,\lambda^*)z>0$, then $(x_0^*,y_0^*)$ is a strict local minimum. 
Specifically, the vector $z$ defined by the set \eqref{eq_Aug8_01} satisfies 
\begin{align}
z^{\text T}\nabla g(x_0^*, y_0^*)=
\begin{bmatrix} z_1 & z_2\end{bmatrix}\begin{bmatrix} -2x_0 \\ -2(1+y_0)(1-\bar r_f^{-2})\end{bmatrix}_{(x_0^*, y_0^*)}=0\label{eq_Nov13_10}
\end{align}
Notice that $x_0^*=0, -2(1+y_0^*)(1-\bar r_f^{-2})\neq 0$. Hence 
 \eqref{eq_Nov13_10} implies that the components $z_1\neq 0$ and $z_2=0$. Obviously it follows from \eqref{eq_Aug28_01} that 
\begin{align}
\begin{bmatrix} z_1 & 0\end{bmatrix}\nabla^2 F(x_0^*,y_0^*,\lambda^*)\begin{bmatrix} z_1 \\ 0\end{bmatrix}=z_1\dfrac{\partial^2 F}{\partial x_0^2}(x_0^*, y_0^*)z_1>0\label{eq_Nov13_08}
\end{align}
Therefore in light of Theorem 3.11 in \citep[p.48]{Avriel}, we can conclude that $(x_0^*, y_0^*)$  is a strict local minimum.  
The same argument can be used to show that $(\hat x_0^*, \hat y_0^*)$ is also a strict local minimum in view of \eqref{eq_Nov13_08}.  Meanwhile a straightforward calculation gives
\[
\Delta v(x_0^*, y_0^*)<\Delta v(\hat x_0^*, \hat y_0^*)
\]
Hence the pair $(x_0^*, y_0^*)$ is a global minimum, and further it is the global minimum due to the uniqueness; see \cite[p.194]{Bertsekas} for the definition of a global minimum.  
This completes the proof of the theorem.

\section{The Hohmann transfer as dynamic optimization problems} \label{sect_2_3}

In this section, 
the orbit transfer problem 
 is formulated  as  
two optimal control problems of a spacecraft in an inverse square law field,  driven by velocity impulses,  
with boundary and interior point constraints. 
The calculus of variations is used to solve the resulting two-point and multi-point boundary-value problems (BCs). 

\subsection{Problem formulation}

Consider the motion of a spacecraft in the inverse square gravitational  field, and the state equation is 
\begin{align}
\dot{\mathbf r}&=\mathbf v\notag\\
\dot{\mathbf v}&=-\dfrac{\mu}{r^3}\mathbf r\label{eq_Jul_4_01a}
\end{align}
where $\mathbf r(t)$ is the spacecraft position vector and $\mathbf v(t)$ is its velocity vector.
The state vector consists of $\mathbf r(t)$ and $\mathbf v(t)$.  
We use \eqref{eq_Jul_4_01a} to describe the state of the transfer orbit, which defines a conic under consideration; see, e.g.,  \cite[Chapter 2]{Curits_book}.  

\begin{problem} \label{problem1}\rm
Given the initial position and velocity vectors of a spacecraft on the initial circular orbit, $\mathbf r(t_0), \mathbf v(t_0)$. 
The terminal time $t_1$ is not specified. 
Let  $t_0^+$ signify just after $t_0$ and $t_1^-$ signify just before $t_1$.\footnote{In mathematical language, $t_0^+$ represents the limit of $t_0$ approached from the right side and $t_1^-$ represents the limit of $t_1$ approached from the left side.}
During the period from $t_0^+$   to $t_1^-$, the state evolves over time according to the equation \eqref{eq_Jul_4_01a}. To guarantee at the time $t_1^+$, the spacecraft enters the finite circular orbit, we impose the following equality constraints on the finial state 
\begin{align}\label{eq_Nov5_01f}
g_{r1}(\mathbf{r}(t_1^+))&=\abs{\mathbf{r}(t_1^+)}-r_f=0\notag\\
g_{v1}(\mathbf{v}(t_1^+))&=\abs{\mathbf{v}(t_1^+)}-v_f
=0
\notag\\
g_{2}(\mathbf{r}(t_1^+), \mathbf{v}(t_1^+))&=
\mathbf r(t_1^+)\cdot \mathbf{v}(t_1^+)=0
\end{align}
where $v_f$ is the orbit velocity of the final circular orbit $v_f=\sqrt{\mu/r_f}$. 
Suppose that there are velocity impulses at time instants $t_0$ and $t_1$
\begin{align}
\mathbf{v}(t_i^+)=\mathbf{v}(t_i^-)+\Delta \mathbf{v}_i, \quad i=0,1
\end{align}
where $\mathbf{v}(t_0^-)=\mathbf{v}(t_0)$. 
The position vector $\mathbf r(t)$ is continuous at these instants.  The optimal control problem is to design $\Delta \mathbf{v}_i$ that minimize the cost functional  
\begin{align}
J=\abs{\Delta\mathbf{v}_0}+\abs{\Delta\mathbf{v}_1} \label{eq_Nov14_01}
\end{align}
subject to the constraints \eqref{eq_Nov5_01f}. 
\end{problem}

In control theory, such an optimal control problem is called impulse control problems in which there are state or control jumps. 
Historically an optimal problem with the cost functional \eqref{eq_Nov14_01} is also referred to  
as the minimum-fuel problem. 
By Problem \ref{problem1},\footnote{
A Matlab script for the Hohmann transfer was given as a numerical example in the free version of a   \texttt{Matlab}-based software \texttt{GPOPS} by an direct method. The constrained \eqref{eq_Nov5_01f} was also used to describe the terminal conditions. We here use the variational method (i.e., indirect method) to optimal control problems}
the orbit transfer problem has been formulated as a dynamic optimization problem instead of a static one considered in Section \ref{sect_2_2}.  The advantage of this formulation is that the coplanar assumption of the transfer orbit to the initial orbit is removed, but the computation complexity follows.  

In the next problem we consider the orbit transfer problem as a dynamic optimization problem with interior point constraints.

\begin{problem}\rm \label{problem2}
Consider a similar situation as in Problem \ref{problem1}.  Let $t_{HT}$ be the Hohmann transfer time. Instead of the unspecified terminal time, here the terminal time instant $t_f>t_{HT}$ is given and the time instant $t_1$ now is an unspecified interior time instant.  The conditions  \eqref{eq_Nov5_01f} become a set of interior boundary conditions.  The optimal control problem is  to design $\Delta \mathbf{v}_i$ that minimize the cost functional \eqref{eq_Nov14_01} subject to the interior boundary conditions \eqref{eq_Nov5_01f}. 
\end{problem}

\subsection{Two-point boundary conditions for Problem \ref{problem1}}\label{sect_2_3_2}

Problem \ref{problem1} is a constrained optimization problem subject to static and dynamic constraints. We use Lagrange multipliers to convert it into an unconstrained one. 
Define the augmented cost functional
\begin{align}
\tilde J:&=\abs{\Delta\mathbf{v}_0}+\abs{\Delta\mathbf{v}_1}\notag\\
&\quad +\mathbf{q}^{\text T}_{r1}\left[\mathbf{r}(t_0^+)-\mathbf{r}(t_0) \right] +\mathbf{q}^{\text T}_{r2}\left[\mathbf{r}(t_1^+)-\mathbf{r}(t_1^-) \right]
\notag\\
&\quad +\mathbf{q}^{\text T}_{v1}\left[\mathbf{v}(t_0^+)-\mathbf{v}(t_0)-\Delta \mathbf{v}_0\right] +\mathbf{q}^{\text T}_{v2}\left[\mathbf{v}(t_1^+)-\mathbf{v}(t_1^-)-\Delta \mathbf{v}_1\right]
\notag\\
&\quad +
\gamma_{r1}g_{r1}(\mathbf r(t_1^+))+\gamma_{v1}g_{v1}(\mathbf v(t_1^+)) +
\gamma_{2}g_{r2}(\mathbf r(t_1^+), \mathbf v(t_1^+))
\notag\\
&\quad+\int_{t_0^+}^{t_1^-}
\mathbf{p}_r^{\text{T}}\left (\mathbf{v}-\dot{\mathbf{r}}\right )+
\mathbf{p}_v^{\text{T}}\left (-\dfrac{\mu}{r^3}\mathbf{r}-\dot{\mathbf{v}}\right ){\rm d}t\notag
\end{align}
where Lagrange multipliers $\mathbf{p}_r, \mathbf{p}_v$ are also called costate vectors; in particular, Lawden \cite{Lawden_book} termed $-\mathbf{p}_v$ the primer vector. 
Introducing Hamiltonian function
\begin{align}
H(\mathbf{r}, \mathbf{v}, \mathbf{p}):=\mathbf{p}_{r}^{\text{T}}\mathbf{v}-\mathbf{p}_{v}^{\text{T}}\dfrac{\mu}{r^3}\mathbf r, \qquad\mathbf{p}=\begin{bmatrix}
\mathbf{p}_{r} & \mathbf{p}_{v}
\end{bmatrix}^{\text T}\label{eq_7_2_1a}
\end{align}
By taking into account of all perturbations, the first variation of the augmented cost functional is 
\begin{align}
\delta \tilde J&=\dfrac{\Delta\mathbf{v}_0^{\text T} }{\abs{\Delta \mathbf{v}_0}}\delta \mathbf{v}_0+
\dfrac{\Delta\mathbf{v}_1^{\text T} }{\abs{\Delta \mathbf{v}_1}}\delta \mathbf{v}_1
\notag\\
&\quad+\mathbf{q}_{r1}^{\text T}\left [d \mathbf r(t_0^{+})-d \mathbf r(t_0^{-})\right]
+\mathbf{q}_{r2}^{\text T}\left[d \mathbf r(t_1^{+})-d \mathbf r(t_1^{-})\right ]
\notag\\
&\quad+\mathbf{q}_{v1}^{\text T}\left [d \mathbf v(t_0^{+})-d \mathbf v(t_0^{-})-\delta \mathbf{v}_0\right]
+\mathbf{q}_{v2}^{\text T}\left[d \mathbf v(t_1^{+})-d \mathbf v(t_1^{-})-\delta \mathbf{v}_1\right ]
\notag\\
&\quad +d \mathbf{r}^{\text T}(t_1^+)\dfrac{\partial g_{r1} (\mathbf{r}(t_1^+))}{\partial \mathbf{r}(t_1^+)}\Big |_*\gamma_{r1}+
d \mathbf{v}^{\text T}(t_1^+)\dfrac{\partial g_{v1} (\mathbf{v}(t_1^+))}{\partial \mathbf{v}(t_1^+)}\Big |_*\gamma_{v1}
\notag\\
&\quad +d \mathbf{r}^{\text T}(t_1^+)\dfrac{\partial g_{2} (\mathbf{r}(t_1^+), \mathbf{v}(t_1^+))}{\partial \mathbf{r}(t_1^+)}\Big |_*\gamma_{2}+
d \mathbf{v}^{\text T}(t_1^+)\dfrac{\partial g_{2} (\mathbf{r}(t_1^+),\mathbf{v}(t_1^+))}{\partial \mathbf{v}(t_1^+)}\Big |_*\gamma_{2}
\notag\\
& \quad
(*)+\left (H_{*}-\mathbf p_{r}^{\text T} \dot{\mathbf r}-\mathbf p_{v}^{\text T} \dot{\mathbf v}\right)\Big |_{t_1^{-*}}\delta t_1\notag
\\
&\quad
(**)+\mathbf p_{r}^{\text T}(t_0^{+})\delta \mathbf r(t_0^{+})-\mathbf p_{r}^{\text T}(t_1^{-*})\delta \mathbf r(t_1^{-*})
+\mathbf p_{v}^{\text T}(t_0^{+})\delta \mathbf v(t_0^{+})
 -\mathbf p_{v}^{\text T}(t_1^{-*})\delta \mathbf v(t_1^{-*})\notag\\
&\quad +\int_{t_0^+}^{t_1^{-*}}\left [\left (\dfrac{\partial H(\mathbf r,\mathbf v, \mathbf p)}{\partial \mathbf r}+\dot{\mathbf p}_{r}(t)\right )^{\text T}_*\delta \mathbf r+\left(\dfrac{\partial H(\mathbf r,\mathbf v, \mathbf p)}{\partial \mathbf v}+\dot{\mathbf p}_{v}(t)\right)^{\text T}_*\delta \mathbf v\right ]\text{d} t\label{eq_Nov15_01}
\end{align}
where we use $d(\cdot)$ to denote the difference between  
the varied path and the optimal path taking into account the differential change in a time instant (i.e., differential in $x$), for example,
\begin{align}
d \mathbf v(t_1^{+})=\mathbf v(t_1^{+})-\mathbf v^*(t_1^{+*})\notag
\end{align}
and $\delta(\cdot)$ is the variation, for example, $\delta \mathbf v(t_1^*)$ is the variation of $\mathbf v$ as an independent variable at $t_1^*$. 
Notice that $d \mathbf r(t_0^{-})=0, d \mathbf v(t_0^{-})=0$ since $t_0$ is fixed.
The parts of $\delta \tilde J$ in \eqref{eq_Nov15_01} marked with asterisks are respectively due to the linear term of 
\begin{align}\notag
\int_{t_1^{-*}}^{t_1^{-*}+\delta t_1}\left(H(\mathbf{r}, \mathbf{v}, \mathbf{p})
-\mathbf{p}_{r}^{\text{T}} \dot{\mathbf{r}} -
\mathbf{p}_{v}^{\text{T}}\dot{\mathbf{v}}\right){\rm d}t
\end{align}
and the first term in the right-hand side  of the following equation 
\begin{align}
\int_{t_0^+}^{t_1^{-*}} \mathbf{p}_{r}^{\text T}\delta \dot{\mathbf{r}}\text{d} t=
\mathbf p_{r}^{\text T}\delta \mathbf r\big |^{t_1^{-*}}_{t_0^+}-\int_{t_0^+}^{t_1^{-*}}
\dot{\mathbf{p}}_{r}^{\text T}\delta\mathbf{r}\text{d} t\notag
\end{align}
obtained by integrating by parts. 
In order to derive boundary conditions, we next use the following relation 
 \begin{align}\label{eq_Nov7_01a}
d \mathbf{v}(t_1^-)=\delta \mathbf{v}(t_1^{-*})+\dot{\mathbf{v}}(t_1^{-*})\delta t_1
 \end{align}
see \citep[Section 3.5]{Bryson_book} and \cite[Section 3.3]{Longuski_book}. 

In view of the necessary condition $\delta \tilde J=0$
and the fundamental lemma, we have the costate equations
\begin{align}
\dot{\mathbf p}_{r}(t)=-\dfrac{\partial H(\mathbf r,\mathbf v, \mathbf p)}{\partial \mathbf r}, \quad 
\dot{\mathbf p}_{v}(t)=-\dfrac{\partial H(\mathbf r,\mathbf v, \mathbf p)}{\partial \mathbf v}\label{eq_Nov7_01}
\end{align}
Based on the definition of the  Hamiltonian function in \eqref{eq_7_2_1a}, the costate equation \eqref{eq_Nov7_01} can be rewritten as 
\begin{align}
\dot{\mathbf{p}}_{r}&=\dfrac{\partial }{\partial \mathbf r}\left(\dfrac{\mu}{r^3}\mathbf r \right)\mathbf{p}_{v}
=-\dfrac{\mu}{r^3}(\dfrac{3}{r^2}\mathbf{r}\mathbf{r}^{\text T}-I_{3\times 3})\mathbf{p}_{v}, \quad 
\dot{\mathbf{p}}_{v}=-\mathbf{p}_{r} \label{eq_July_4_02}
\end{align}
where $I_{3\times 3}$ is the $3\times 3$ identity matrix.

Using \eqref{eq_Nov7_01a} and regrouping terms in \eqref{eq_Nov15_01} yields the following terms or equalities for $\Delta \mathbf{v}_1, \Delta \mathbf{v}_2, d \mathbf{v}$  \begin{align}
\left(\dfrac{\Delta \mathbf{v}_0}{\abs{\Delta \mathbf{v}_0}}-\mathbf{q}_{v1}\right)^{\text T}\Delta \mathbf v_0\notag\\
\left(\dfrac{\Delta \mathbf{v}_1}{\abs{\Delta \mathbf{v}_1}}-\mathbf{q}_{v2}\right)^{\text T}\Delta \mathbf v_1\notag\\
\mathbf{q}_{v1}^{\text T} d \mathbf{v}(t_0^+)
+\mathbf{p}_{v}^{\text T}(t_0^+) \delta \mathbf{v}(t_0^{+})=\left(\mathbf{q}_{v1}^{\text T} +\mathbf{p}_{v}^{\text T}(t_0^+) \right)d \mathbf v(t_0^{+}) \notag\\
-\mathbf{q}_{v2}^{\text T} d \mathbf v(t_1^{-}) 
-\mathbf{p}_{v}^{\text T}(t_1^-)\left(\delta \mathbf{v}(t_1^{-*})+\dot{\mathbf{v}}(t_1^{-*})\delta t_1\right)=-\left(\mathbf{q}_{v2}^{\text T} +\mathbf{p}_{v}^{\text T}(t_1^-)\right)d \mathbf v(t_1^{-}) \notag\\
\left(\mathbf{q}_{v2}^{\text T} 
+\gamma_{v1}\dfrac{\partial g_{v1} (\mathbf{v}(t_1^+))}{\partial \mathbf{v}^{\text T}(t_1^+)}\Big |_*+
\gamma_{2}\dfrac{\partial g_{2} (\mathbf{r}(t_1^+),\mathbf{v}(t_1^+))}{\partial \mathbf{v}^{\text T}(t_1^+)}\Big |_*
\right)
d \mathbf v(t_1^{+}) \label{eq_Nov15_02}
 \end{align}
 As usual, in order to assure $\delta\tilde J=0$, we choose Lagrange multipliers to make the coefficients of 
 $\Delta \mathbf{v}_0$, $\Delta \mathbf{v}_1$,  $d \mathbf v(t_0^{+})$, $d \mathbf v(t_1^{-})$, $d \mathbf v(t_1^{+})$ in \eqref{eq_Nov15_02} vanish respectively
 \begin{align}
\mathbf{q}_{v1}-\dfrac{\Delta \mathbf{v}_0}{\abs{\Delta \mathbf{v}_0}}=0, \quad
 \mathbf{q}_{v2}-\dfrac{\Delta \mathbf{v}_1}{\abs{\Delta \mathbf{v}_1}}& =0\notag\\\mathbf{p}_{v} (t_0^+)+\mathbf{q}_{v1}=0, \quad
\mathbf p_{v}(t_1^-)+\mathbf q_{v2}&=0\notag\\
\mathbf{q}_{v2} 
+\gamma_{v1}\dfrac{\partial g_{v1} (\mathbf{v}(t_1^+))}{\partial \mathbf{v} (t_1^+)}\Big |_*+
\gamma_{2}\dfrac{\partial g_{2} (\mathbf{r}(t_1^+),\mathbf{v}(t_1^+))}{\partial \mathbf{v} (t_1^+)}\Big |_* &=0 \label{eq_Nov7_05}
\end{align} 
Applying the similar argument as above for $\mathbf p_r$ gives
\begin{align}
\mathbf{p}_{r} (t_0^+)+\mathbf{q}_{r1} =0, \quad 
\mathbf p_{r}(t_1^-)+\mathbf q_{r2}&=0\notag\\
\mathbf{q}_{r2} 
+\gamma_{r1}\dfrac{\partial g_{r1} (\mathbf{r}(t_1^+))}{\partial \mathbf{r} (t_1^+)}\Big |_*+
\gamma_{2}\dfrac{\partial g_{2} (\mathbf{r}(t_1^+),\mathbf{v}(t_1^+))}{\partial \mathbf{r} (t_1^+)}\Big |_* &=0 \label{eq_Nov7_06}
\end{align}
We now choose $H(t_1^-)$  to cause the coefficient of $\delta t_1$ in \eqref{eq_Nov15_01} to vanish
\begin{align}
H(t_1^-)&=\mathbf{p}_{r}^{\text{T}}(t_1^-)\mathbf{v}(t_1^-)-\mathbf{p}_{v}^{\text{T}}(t_1^-)\dfrac{\mu}{r^3(t_1^-)}\mathbf{r}(t_1^-)=0\notag
\end{align}

Finally, by solving the first and second component equations of the last vector equation 
in \eqref{eq_Nov7_05}, we obtain the Lagrangian multipliers $\gamma_{v1}, \gamma_{2}$, and then substituting them into the first component equations of the last vector equation 
in \eqref{eq_Nov7_06} yields the Lagrangian multiplier $\gamma_{r1}$. With these  Lagrangian multipliers and rearranging the second and third component equations of the last vector equation in \eqref{eq_Nov7_06}, we obtain a boundary value equation denoted by 
\[
\mathbf g\left(\mathbf{r}(t_1^+), \mathbf{v}(t_1^+); \mathbf{p}_{r}(t_1^-), \mathbf{p}_{v}(t_1^-)\right)=0
\] 

\begin{bcs} \label{table_Nov7_01}
\begin{center} 
\centering{\small Two-point boundary conditions for Problem \ref{problem1} \vspace*{0.2cm}}\\
{\blue
\fbox{
\begin{minipage}[H]{10cm} \vspace*{-.3cm}
\black\small
\begin{align}
\begin{array}{ll}
&(1)~
\mathbf{r}(t_0)-\mathbf{r}_0=0, \quad \mathbf{v}(t_0^+)-\mathbf{v}(t_0^-)-\Delta \mathbf v_0=0\notag
\vspace*{0.1cm}
\\
&(2)~
 \mathbf p_{Mv}(t_0^+)+
\dfrac{\Delta \mathbf{v}_0}{\abs{\Delta \mathbf{v}_0}} =0, \quad 
 \mathbf p_{Mv}(t_1^-)+
\dfrac{\Delta \mathbf{v}_1}{\abs{\Delta \mathbf{v}_1}} =0
\vspace*{0.1cm}
\\
&(3)~ 
H(t_1^-)=0 \\
\vspace*{0.1cm}
&(4)\begin{cases}
\abs{\mathbf{r}(t_1^+)}-r_f=0\notag\\
\abs{\mathbf{v}(t_1^+)}-v_f
=0
\notag\\
\mathbf r(t_1^+)\cdot \mathbf{v}(t_1^+)=0
\end{cases}
\vspace*{0.1cm}
\\
&(5)
~\mathbf g\left (\mathbf{r}(t_1^+), \mathbf{v}(t_1^+); \mathbf{p}_{r}(t_1^-), \mathbf{p}_{v}(t_1^-)\right)=0\\
\end{array}
\notag 
\end{align}
\end{minipage}}}
\end{center}
\end{bcs}
In summary, a complete list including 19 BCs is shown in List \ref{table_Nov7_01} where
$
\mathbf{r}(t_1^+)=\mathbf{r}(t_1^-), \mathbf{v}(t_1^+)=\mathbf{v}(t_1^-)+\Delta \mathbf v_1. 
$

\subsection{Multi-point boundary conditions for Problem \ref{problem2}}
Define the augmented cost functional 
\begin{align}
\tilde J:&=\abs{\Delta\mathbf{v}_0}+\abs{\Delta\mathbf{v}_1}\notag\\
&\quad +\mathbf{q}^{\text T}_{r1}\left[\mathbf{r}(t_0^+)-\mathbf{r}(t_0) \right] +\mathbf{q}^{\text T}_{r2}\left[\mathbf{r}(t_1^+)-\mathbf{r}(t_1^-) \right]
\notag\\
&\quad +\mathbf{q}^{\text T}_{v1}\left[\mathbf{v}(t_0^+)-\mathbf{v}(t_0)-\Delta \mathbf{v}_0\right] +\mathbf{q}^{\text T}_{v2}\left[\mathbf{v}(t_1^+)-\mathbf{v}(t_1^-)-\Delta \mathbf{v}_1\right]
\notag\\
&\quad +
\gamma_{r1}g_{r1}(\mathbf r(t_1^+))+\gamma_{v1}g_{v1}(\mathbf v(t_1^+)) +
\gamma_{2}g_{r2}(\mathbf r(t_1^+), \mathbf v(t_1^+))
\notag\\
&\quad+\int_{t_0^+}^{t_1^-}
\mathbf{p}_r^{\text{T}}\left (\mathbf{v}-\dot{\mathbf{r}}\right )+
\mathbf{p}_v^{\text{T}}\left (-\dfrac{\mu}{r^3}\mathbf{r}-\dot{\mathbf{v}}\right ){\rm d}t+\int_{t_1^+}^{t_f}
\mathbf{p}_r^{\text{T}}\left (\mathbf{v}-\dot{\mathbf{r}}\right )+
\mathbf{p}_v^{\text{T}}\left (-\dfrac{\mu}{r^3}\mathbf{r}-\dot{\mathbf{v}}\right ){\rm d}t\notag
\end{align}
We introduce the Hamiltonian functions for two time sub-intervals $[t_0^+, t_1^-] \cup [t_1^+, t_f]$
\begin{align}
H_i(\mathbf{r}, \mathbf{v}, \mathbf{p}_i):=\mathbf{p}_{ri}^{\text{T}}\mathbf{v}-\mathbf{p}_{vi}^{\text{T}}\dfrac{\mu}{r^3}\mathbf r, \qquad\mathbf{p}_i=\begin{bmatrix}
\mathbf{p}_{ri} & \mathbf{p}_{vi}
\end{bmatrix}^{\text T}, 
\qquad i=1,2\label{eq_7_2_1}
\end{align}

A tedious and similar argument as used in Problem \ref{problem1} can be applied to Problem \ref{problem2} to derive the costate equations and boundary conditions. 
The two-phase costate equations are given by
\begin{align}
\dot{\mathbf p}_{ri}(t)&=-\dfrac{\partial H_i(\mathbf r,\mathbf v, \mathbf p_i)}{\partial \mathbf r}, \quad
\dot{\mathbf p}_{vi}(t)=-\dfrac{\partial H_i(\mathbf r,\mathbf v, \mathbf p_i)}{\partial \mathbf v}, \quad i=1,2\notag
\end{align}
A complete list for boundary conditions is  given  in List \ref{table_Nov7_02}. 

\begin{bcs} \label{table_Nov7_02}
\begin{center} 
\centering{\small 31 boundary conditions for Problem \ref{problem2} \vspace*{0.2cm}}\\
{\blue
\fbox{
\begin{minipage}[H]{10cm} \vspace*{-.3cm}
\black\small
\begin{align}
\begin{array}{ll}
&(1)\begin{cases}
\mathbf{r}(t_0)-\mathbf{r}_0=0, \quad \mathbf{v}(t_0^-)-\mathbf{v}(t_0^+)-\Delta \mathbf v_0=0\notag\\
\mathbf{r}(t_1^-)-\mathbf{r}(t_1^+)=0, \quad \mathbf{v}(t_1^-)-\mathbf{v}(t_1^+)-\Delta \mathbf v_1=0\notag\\
\end{cases}
\vspace*{0.1cm}
\\
&(2)\begin{cases}
 \mathbf p_{v1}(t_0^+)+
\dfrac{\Delta \mathbf{v}_0}{\abs{\Delta \mathbf{v}_0}} =0, \quad 
 \mathbf p_{v1}(t_1^-)+
\dfrac{\Delta \mathbf{v}_1}{\abs{\Delta \mathbf{v}_1}} =0\\
\mathbf p_{v2}(t_f)=0, \quad
\mathbf p_{r2}(t_f)=0
\end{cases}
\vspace*{0.1cm}
\\
&(3)~ 
H_1(t_1^-)-H_2(t_1^+)=0 
\text{~or~}-\mathbf p_{r1}^{\rm T}(t_1^{-})\Delta \mathbf v_1=0\\
\vspace*{0.1cm}
&(4)\begin{cases}
\abs{\mathbf{r}(t_1^+)}-r_f=0\notag\\
\abs{\mathbf{v}(t_1^+)}-v_f
=0
\notag\\
\mathbf r(t_1^+)\cdot \mathbf{v}(t_1^+)=0
\end{cases}
\vspace*{0.1cm}
\\
&(5)
~\mathbf g\left (\mathbf{r}(t_1^+), \mathbf{v}(t_1^+); \mathbf{p}_{r1}(t_1^-), 
\mathbf{p}_{r1}(t_1^+), \mathbf{p}_{v1}(t_1^-), 
\mathbf{p}_{v1}(t_1^+) \right)=0
\vspace*{0.1cm}
\\
\end{array}
\notag 
\end{align}
\end{minipage}}}
\end{center}
\end{bcs}

\section{Numerical Examples} \label{sect_2_3_4}
\begin{figure}[htb!]
 \hspace{-1cm}
 \begin{minipage}{8cm}
\centering
\includegraphics[scale=.6]{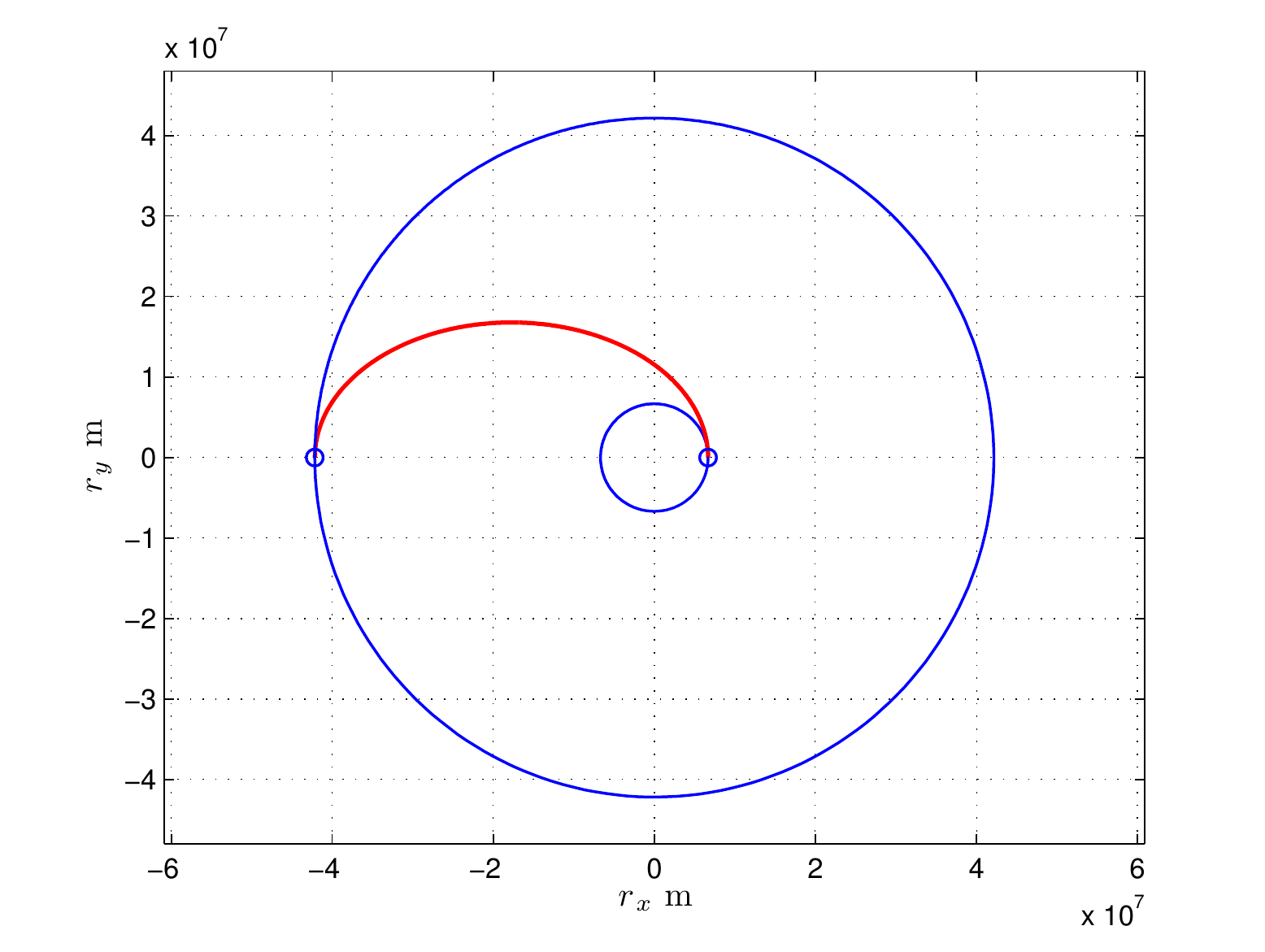}
\end{minipage}
\qquad
 \begin{minipage}{8cm}
\centering
\includegraphics[scale=.6]{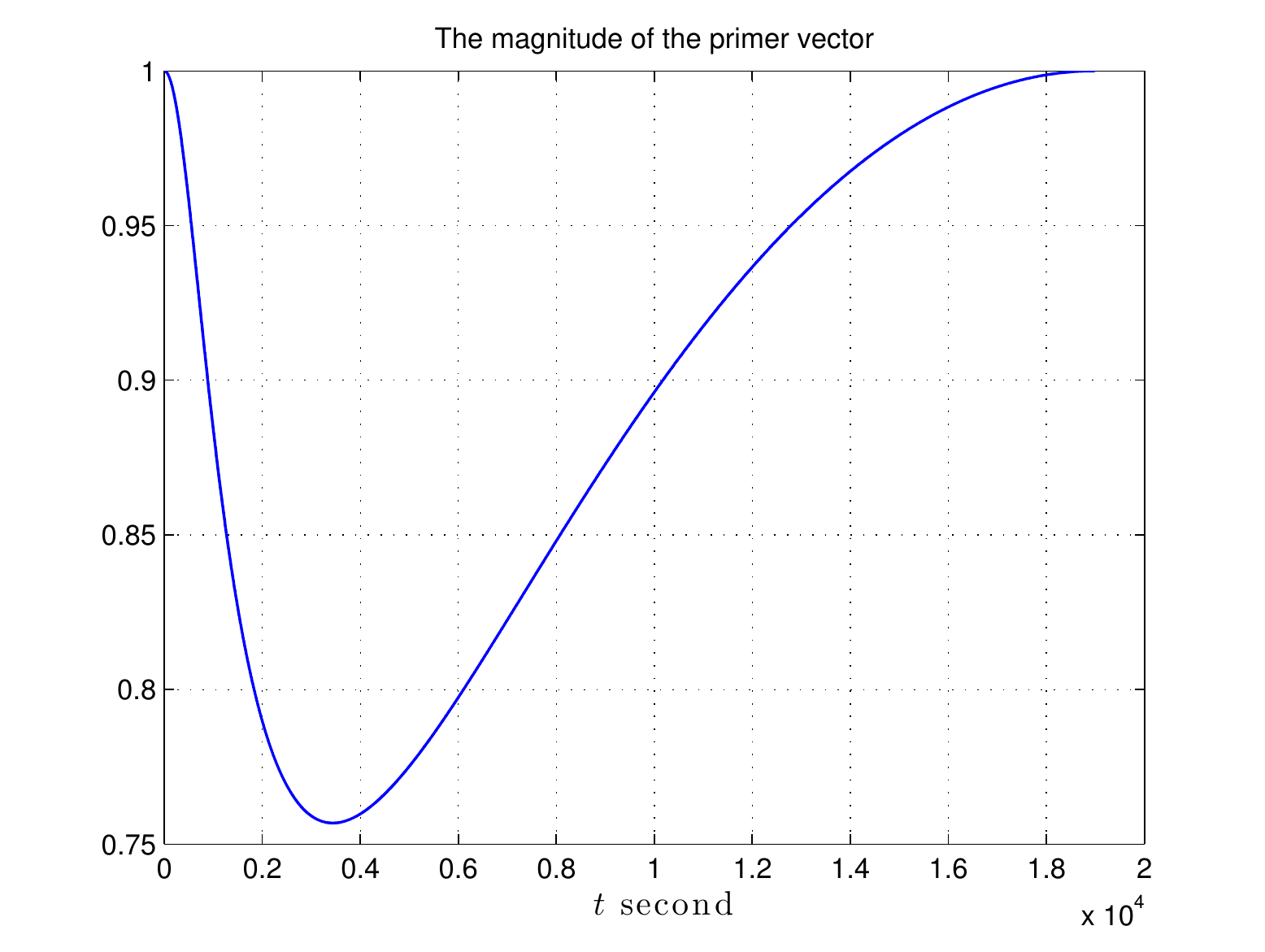}
\end{minipage}
\caption{The Hohmann transfer and the magnitude of the primer vector}\label{fig_exam3_2g}
 \end{figure}

In order to use \texttt{Matlab} solver \texttt{bvp4c} or \texttt{bvp5c} to solve two-point and multi-point boundary value problems with unspecified switching time instants in Section \ref{sect_2_3}, two time changes must be introduced. For the time change,  we refer to, e.g., \cite[Appendix A]{Longuski_book} and \cite[Section 3]{Zefran} for details.    
The solver \texttt{bvp4c} or \texttt{bvp5c} accepts boundary value problems with unknown parameters; see \cite{Shampine_book} and \cite{Kierzenka_thesis}. 

 \begin{exam}\label{exam_2_3_2} \rm


This example in \cite[p.25]{Marec_book} is used to illustrate that the Hohmann transfer is the solution to Problem \ref{problem1}. 
The altitude of the initial circular orbit is 300km, and  
the desired final orbit is geostationary, that is, its radius is 42164km. 
By the formulas of the Hohmann transfer, the magnitudes of two velocity impulses are given by
\begin{align}
\Delta v_1 = 2.425726280326563e+03, \quad 
\Delta v_2 = 1.466822833675619e+03 \quad \text{m/s}\label{eq_Oct27_01}
\end{align}
Setting the tolerance of {\rm\texttt{bvp4c}} equal to {\rm\texttt{1e-6}}, we have the solution to the minimum-fuel problem \ref{problem1} with an unspecified terminal time given by {\rm\texttt{bvp4c}}
\begin{verbatim}
dv1 = 1.0e+03*[0.000000000000804;  2.425726280326426;  0]
dv2 = 1.0e+03*[0.000000000000209;  -1.466822833675464;  0]
\end{verbatim}   
Compared with \eqref{eq_Oct27_01}, the accuracy of the numerical solution is found to be satisfactory since only the last three digits of fifteen digits after decimal place are different. It is desirable to speed up the computational convergence by scaling the state variables though {\rm\texttt{bvp4c}} already has a scale procedure. The computation time is about
180 seconds  on  Intel Core i5 (2.4 GHz, 2 GB). Figure \ref{fig_exam3_2g}
shows the Hohmann transfer and the magnitude of the primer vector.
The solution provided by {\rm\texttt{bvp4c}} depends upon initial values.  Figure \ref{fig_exam3_2s}
shows the transfer orbit and the magnitude of the primer vector corresponding to the local minimum given by \eqref{eq_Nov13_09} in the proof of Theorem \ref{thm01}.

\begin{figure}[h!]
 \hspace{-1cm}
 \begin{minipage}{8cm}
\centering
\includegraphics[scale=.6]{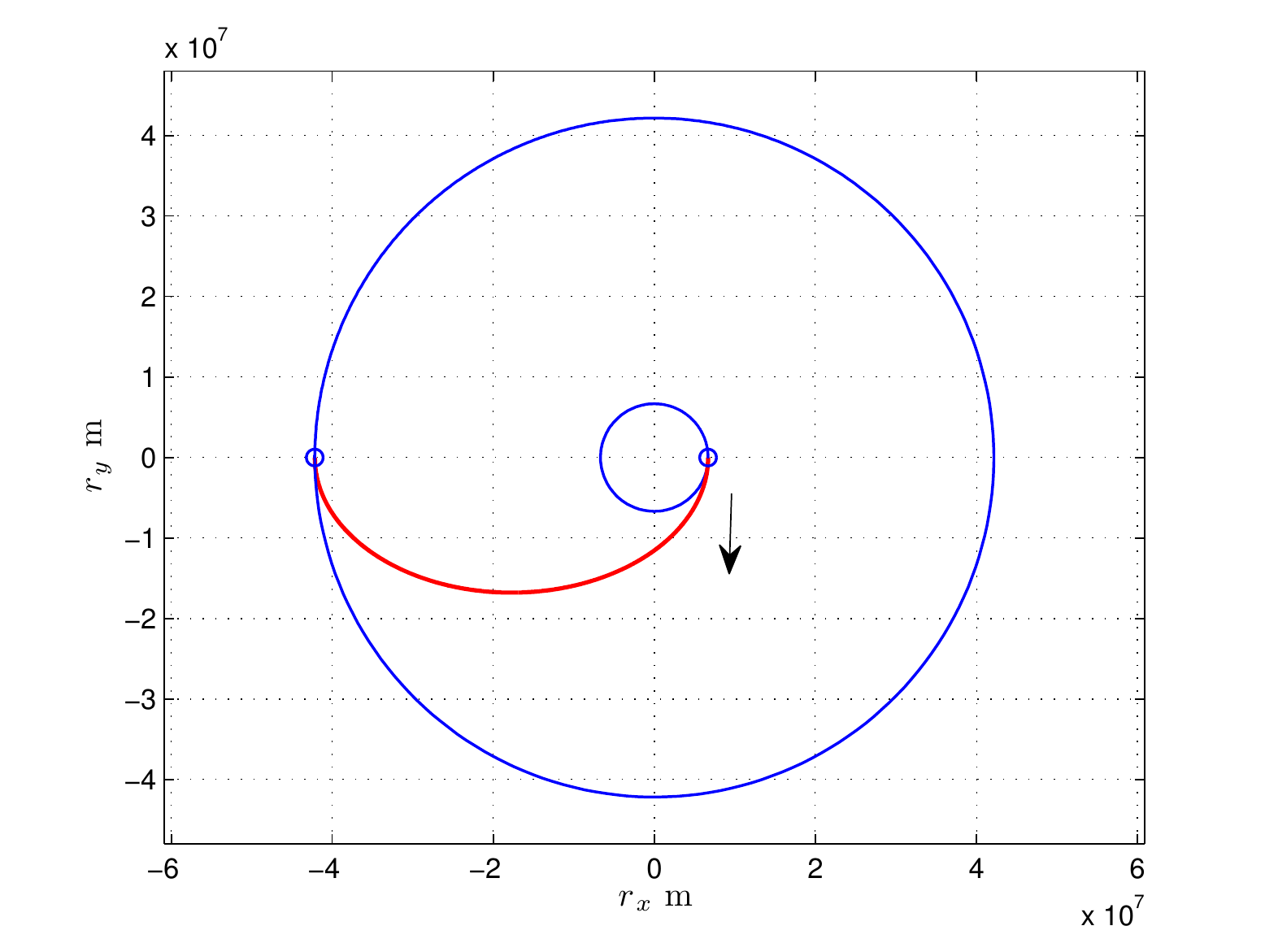}
\end{minipage}
\qquad
 \begin{minipage}{8cm}
\centering
\includegraphics[scale=.6]{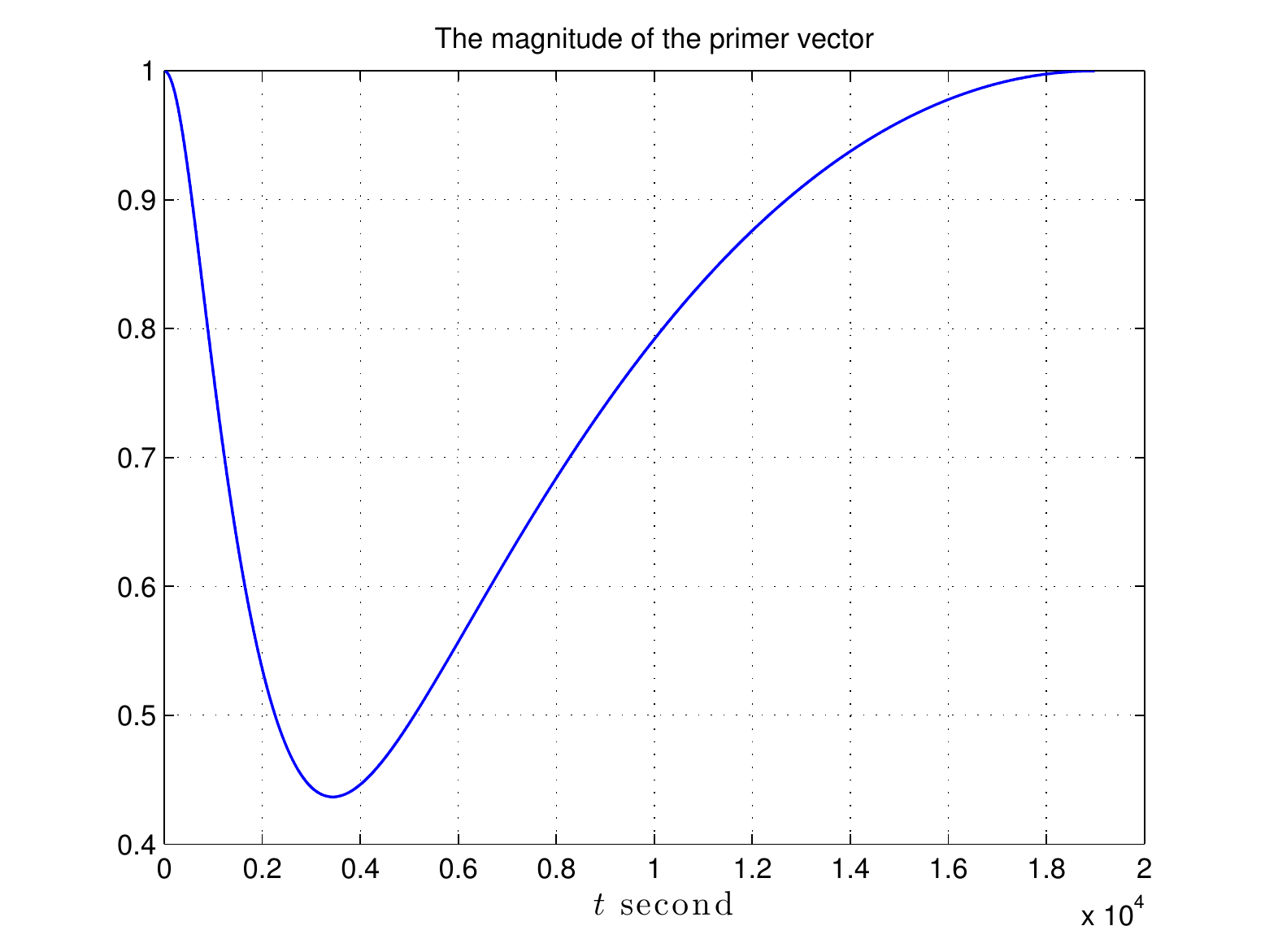}
\end{minipage}
\caption{The orbit transfer and the magnitude of the primer vector for the local minimum}\label{fig_exam3_2s}
 \end{figure}

\end{exam}

\begin{figure}[htb!]
  \hspace{-1cm}
 \begin{minipage}{8cm}
\centering
\includegraphics[scale=.6]{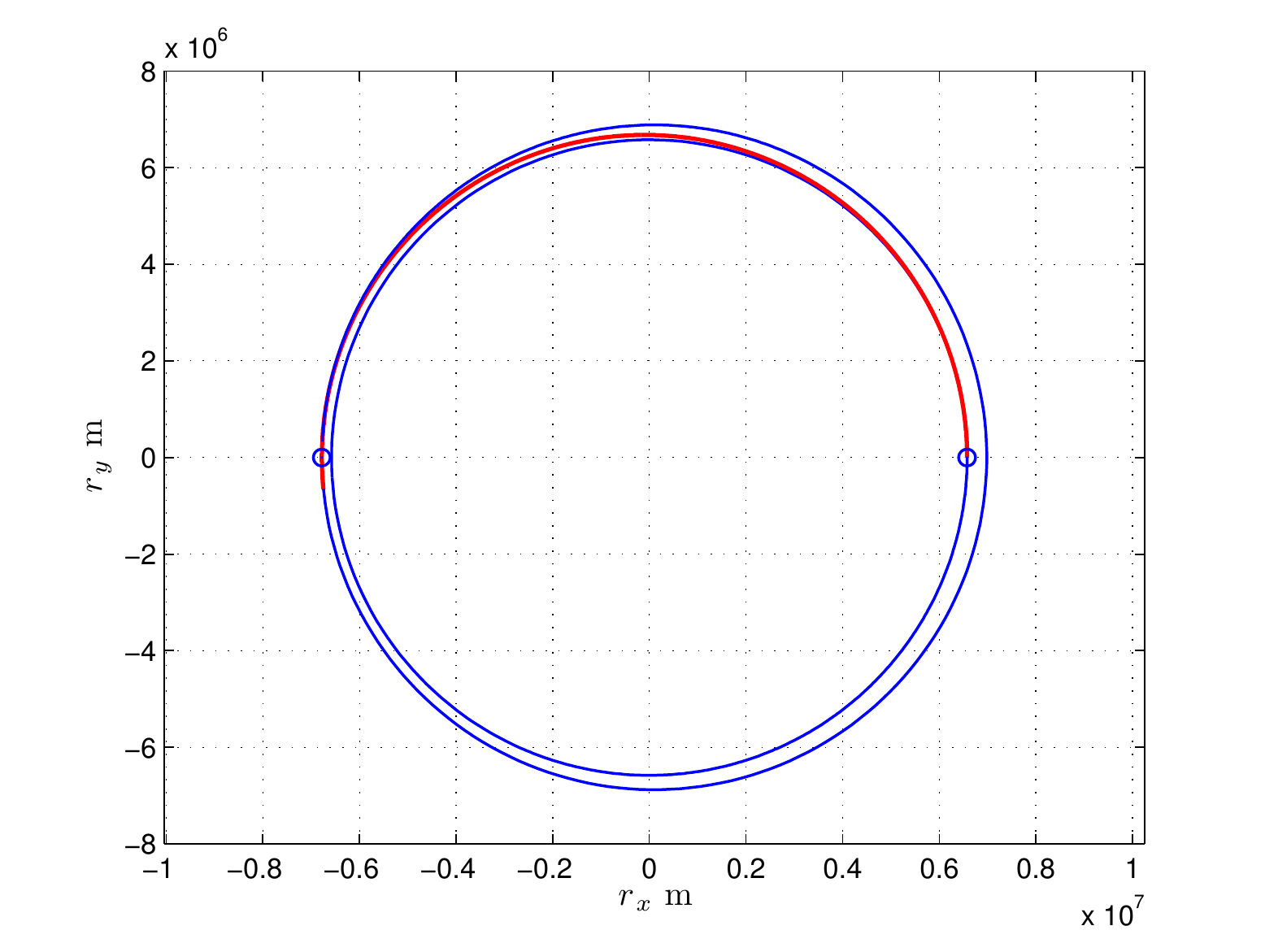}
\end{minipage}
\qquad
 \begin{minipage}{8cm}
\centering
\includegraphics[scale=.6]{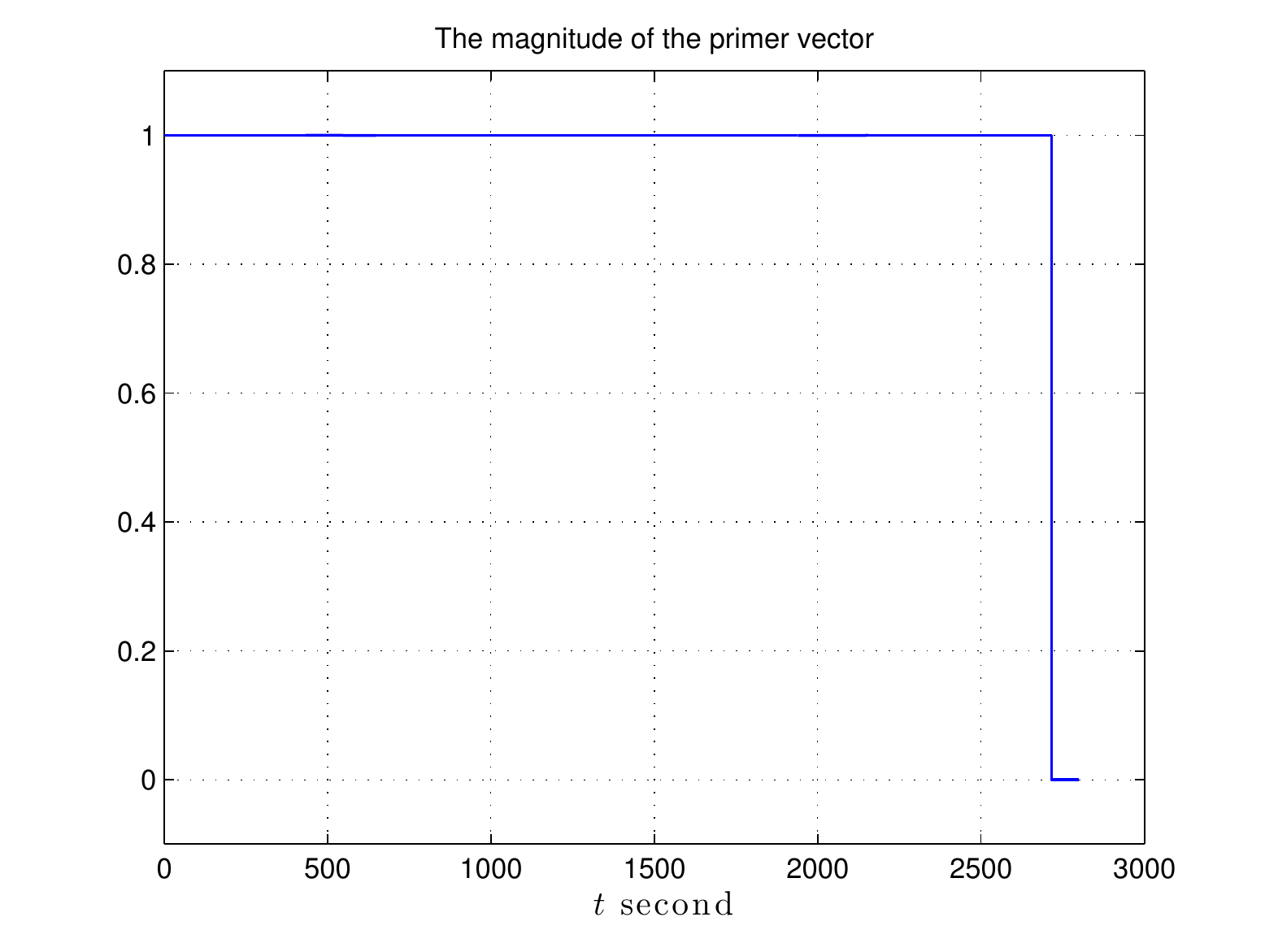}
\end{minipage}
\caption{The Hohmann transfer and the magnitude of the primer vector}\label{fig_exam03_1}
\end{figure}

\begin{exam} \rm


Consider the Hohmann transfer as the multi-point boundary value problem (Problem \ref{problem2}).
The altitudes of the initial and final circular orbits are 200km and 400km respectively.
By the formulas of the Hohmann transfer, the magnitudes of two velocity impulses are given by
\begin{align}
\Delta v_1 = 58.064987253967857, \quad 
\Delta v_2 = 57.631827424189602
\end{align}
The Hohmann transfer time $t_{HT}=2.715594949192177e+03$ seconds, 
hence we choose the terminal time $t_f=2800$ seconds.
The constants, the initial values of the unknown parameters, and the solver and its tolerance are specified in Table \ref{table_3_02}.
\begin{center}
\begin{threeparttable}
\small 
\caption{Constants and initial values}\label{table_3_02}
\begin{tabular}{c|c c c | c c c}
\toprule
Constants &\multicolumn{3}{c}{Gravitational constant $\mu =3.986e+14$} &
\multicolumn{3}{c}{Earth's radius $Re=6378145$}\\
\hline
\multirow{2}{*}{The initial values of the state} & \multicolumn{3}{c}{$\mathbf r_0$ m} \vline & \multicolumn{3}{c}{$\mathbf v_0$ m/s}\\
\cline{2-7}
& 6578145& \qquad 0 
 & 0 &0 & \qquad 7.7843e+03& 0\\
\hline
\multirow{2}{*}{The initial values of the costate} & \multicolumn{3}{c}{$\mathbf p_v$} \vline & \multicolumn{3}{c}{$\mathbf p_r$}\\
\cline{2-7}
& -0.0012 &\qquad 0 & 0 &  0 &   \qquad -0.9& 0\\
\hline
\multirow{2}{*}{The initial values of the velocity impulses} & \multicolumn{3}{c}{$\Delta \mathbf v_1$}  \vline & \multicolumn{3}{c}{$\Delta \mathbf v_2$}\\
\cline{2-7}
& 0 & \qquad 102 & 0 & 0 &  \qquad 102 & 0\\
\hline
The initial value of the scaled time &\multicolumn{6}{c}{~~~0.9} \\
\hline
\multicolumn{4}{c}{Solver {\rm\texttt{bvp4c}}} &
\multicolumn{3}{c}{{\rm\texttt{Tolerance=1e-7}}}\\
\bottomrule
\end{tabular}
\end{threeparttable}
\end{center}
\medskip

By trial and error,  
this example is successfully solved by using
 {\rm \texttt{bvp4c}}.  We obtain the following solution message 
\begin{verbatim}
The solution was obtained on a mesh of 11505 points.
The maximum residual is  4.887e-08. 
There were 2.2369e+06 calls to the ODE function. 
There were 1442 calls to the BC function. 
Elapsed time is 406.320776 seconds.
The first velocity impulse vector dv1 = [0.000000000000001,  58.064987253970472, 0]
The second velocity impulse vector dv2 = [0.000000000000924, -57.631827424187151, 0]
The scaled instant of the second velocity impulse 0.969855338997211
The time instant of the second velocity impulse 2.715594949192190e+03
The maximal error of boundary conditions 4.263256e-13
\end{verbatim}

\begin{figure}[htb]
  \hspace{-1cm}
 \begin{minipage}{8cm}
\centering
\includegraphics[scale=.6]{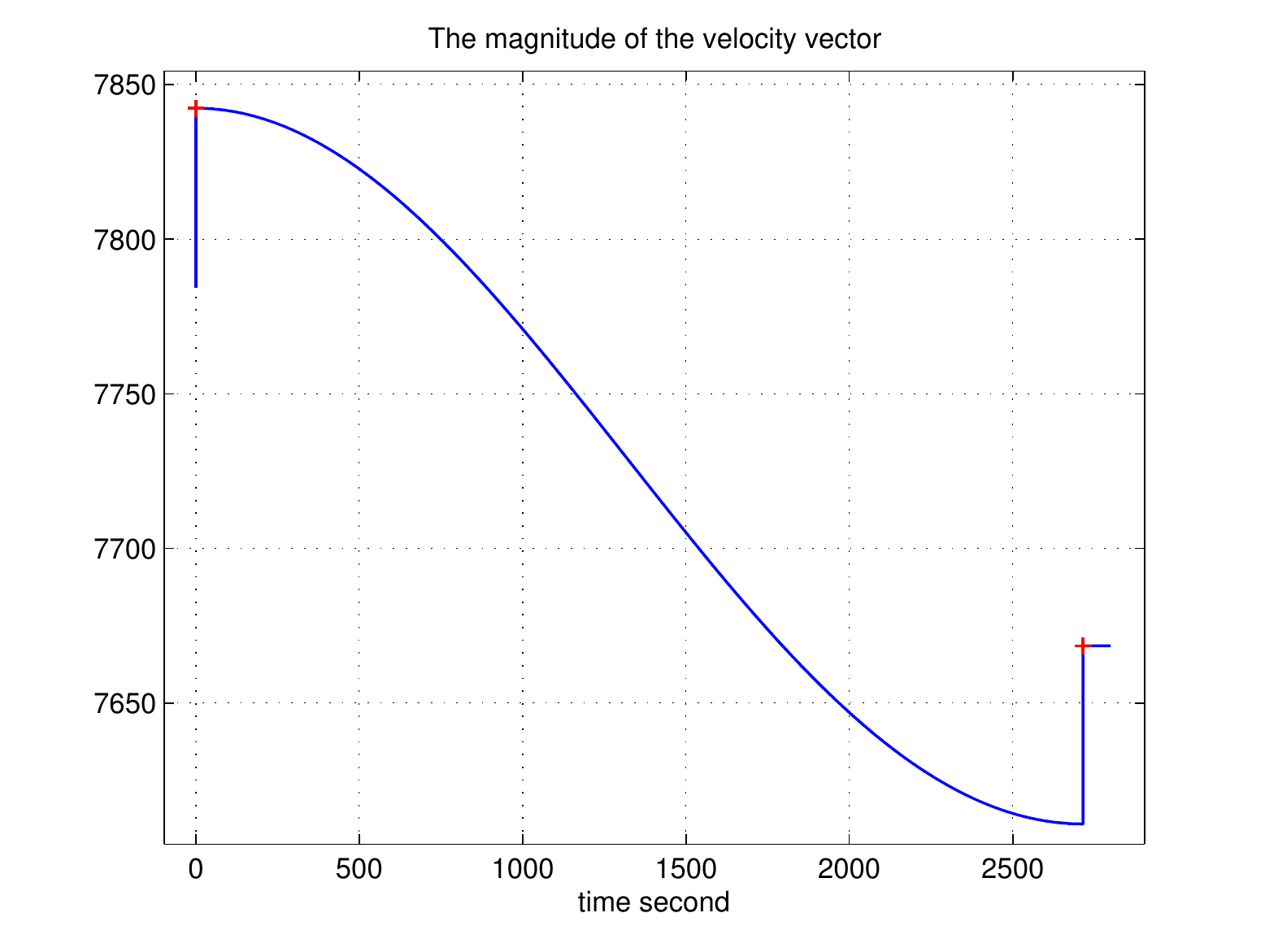}
\end{minipage}
\qquad
 \begin{minipage}{8cm}
\centering
\includegraphics[scale=.6]{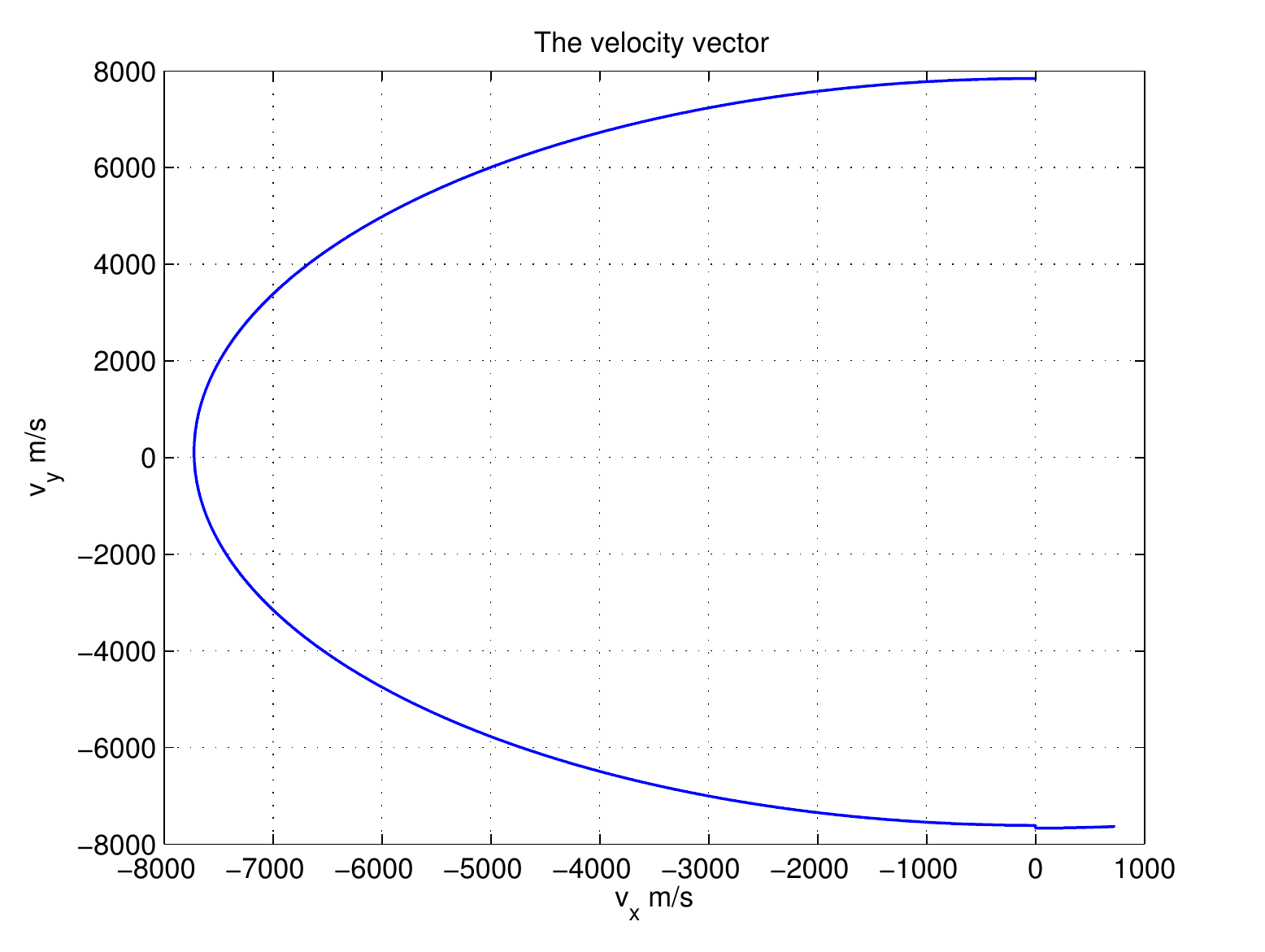}
\end{minipage}
\caption{The magnitude of the velocity and the velocity vector on the $v_x-v_y$ plane}\label{fig_exam03_2}
\end{figure}
Figure \ref{fig_exam03_1}
shows the Hohmann transfer and the magnitude of the primer vector.
The magnitude of the velocity and the velocity vector on the $v_x-v_y$ plane are shown in Figure \ref{fig_exam03_2}, where the symbol $+$ corresponds to the time instant at which the velocity impulse occurs. 

We now assume the initial values of the velocity impulses 
\begin{verbatim}
dv10=[0, 45, 0],  dv20=[0, -60, 0]  
\end{verbatim}
In stead of {\rm \texttt{bvp4c}}, we use the other solver {\rm \texttt{bvp5c}}. Setting  {\rm\texttt{Tolerance=1e-6}}, 
the solution corresponding to the local minimum is found. 
\begin{verbatim}
The solution was obtained on a mesh of 2859 points.
The maximum error is  5.291e-14. 
There were 809723 calls to the ODE function. 
There were 1382 calls to the BC function. 
Elapsed time is 190.920474 seconds.
The first velocity impulse vector  dv1 = 1.0e+4*[0.000000000000001, -1.562657038966310, 0]
The second velocity impulse vector  dv2 = 1.0e+4*[0.000000000000011, -1.527946697280544, 0]
The scaled instant of the second velocity impulse 0.969855338997205
The time instant of the second velocity impulse 2.715594949192175e+03
The maximal error of boundary conditions 4.403455e-10
\end{verbatim}
 \begin{figure}[htb]
 \hspace{-1cm}
 \begin{minipage}{8cm}
\centering
\includegraphics[scale=.6]{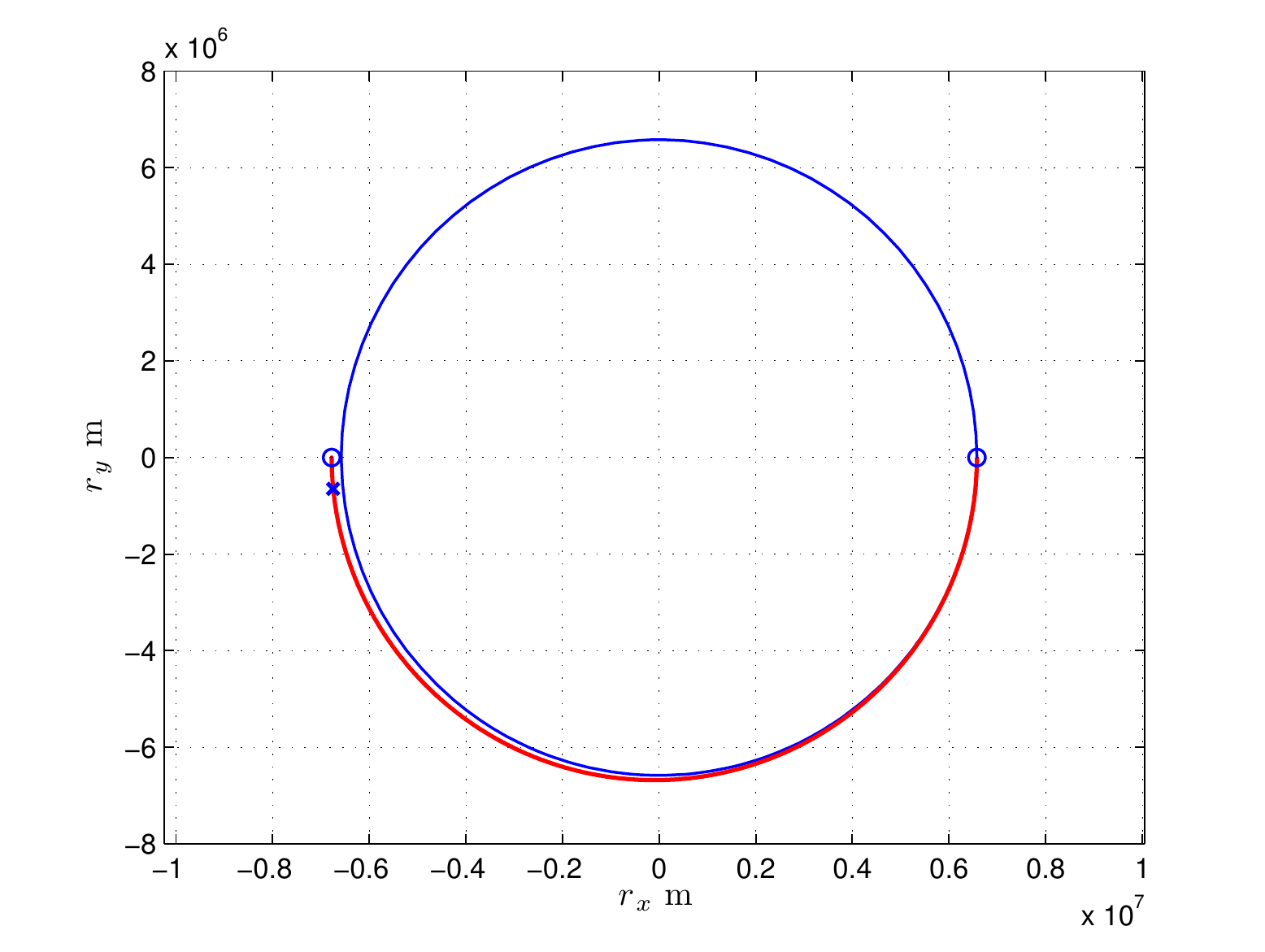}
\end{minipage}
\qquad
 \begin{minipage}{8cm}
\centering
\includegraphics[scale=.6]{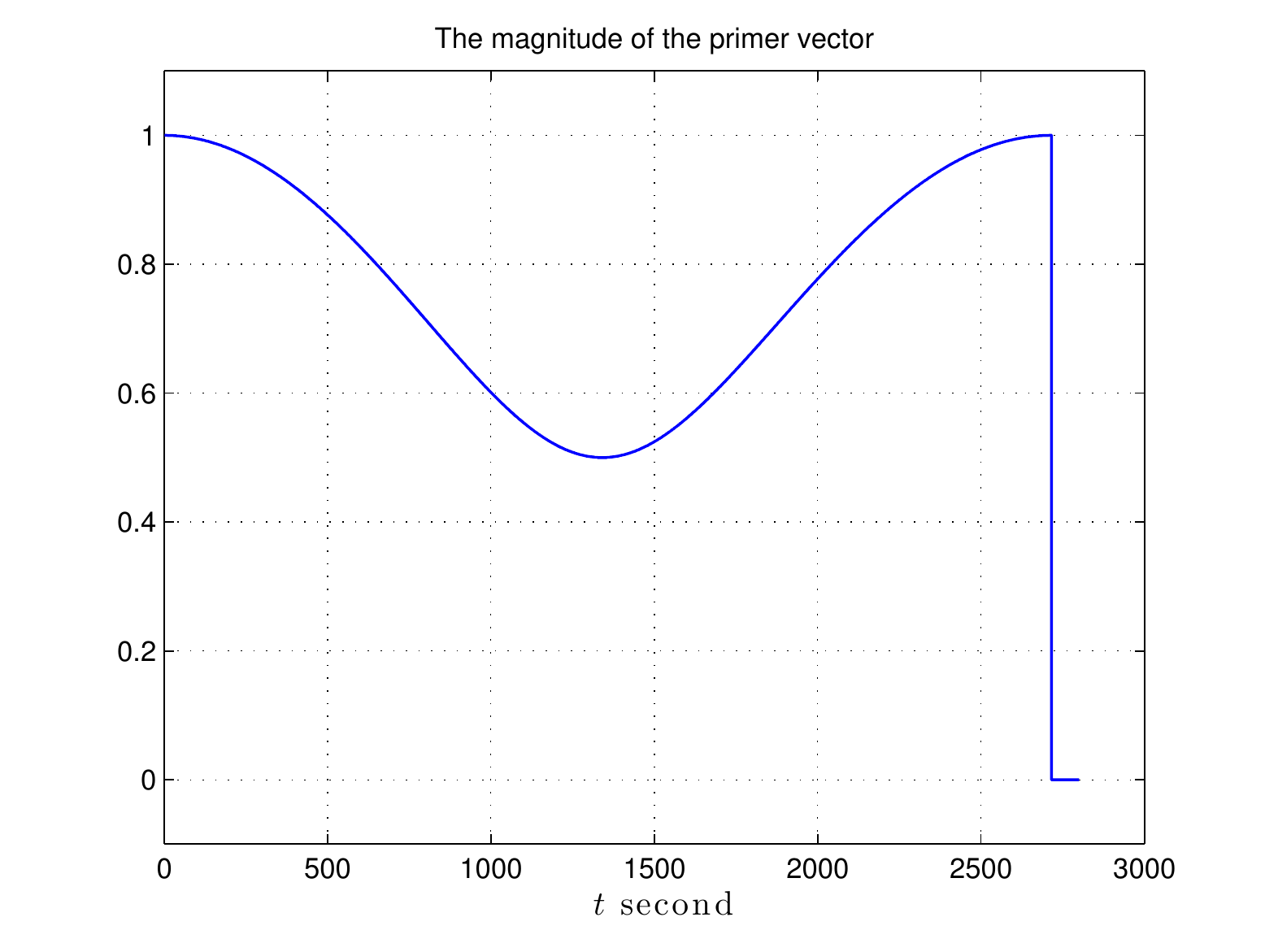}
\end{minipage}
\caption{The orbit transfer and the magnitude of the primer vector corresponding to the local minimum}\label{fig_exam3_2}
\end{figure}
Figure \ref{fig_exam3_2} shows the orbit transfer and the magnitude of the primer vector corresponding to the local minimum, where
the spacecraft starts from the initial point on the initial orbit, moves clockwise  along the transfer orbit until the second pulse point, and then travels  counterclockwise along the final circle orbit until the terminal point marked with the symbol {\rm $\times$}.
\end{exam}

\section{Conclusion}

In this paper, by a static constrained optimization, we study the global optimality of the Hohmann transfer using a nonlinear programming method. Specifically, an inequality presented by  Marec in \citep[pp. 21-32]{Marec_book} is used to define an inequality constraint, then we formulate the Hohmann transfer problem as a constrained nonlinear programming problem. A natural application of the well-known results in nonlinear programming such as the Kuhn-Tucker theorem clearly shows the the global optimality of the Hohmann transfer. In the second part of the paper, we introduce two optimal control problems with two-point and multi-point boundary value constraints respectively. With the help of \texttt{Matlab} solver \texttt{bvp4c}, the Hohmann transfer is solved successfully. 

\section*{Acknowledgments}
The first author is supported by the National Natural Science Foundation of China (no. 61374084).

\addcontentsline{toc}{chapter}{References}

\bibliography{XZY_01_arXiv}

\end{document}

\end{document}